\documentclass[aps,prl,twocolumn,showpacs,superscriptaddress,floatfix]{revtex4-1}
\usepackage{graphicx,graphics}
\usepackage{dcolumn}
\usepackage{amsmath,amssymb,amsfonts}
\usepackage{latexsym,verbatim}
\usepackage{bm}
\usepackage{color}
\usepackage[breaklinks=true,colorlinks,citecolor=blue,linkcolor=blue,urlcolor=blue]{hyperref}
\usepackage{tabularx}

\def\be{\begin{eqnarray}}
\def\ee{\end{eqnarray}}

\begin{document}
\title{Fractional quantum Hall effect in strained graphene: stability of Laughlin states in disordered (pseudo)magnetic fields}
\author{Andrey A. Bagrov}
\email{a.bagrov@science.ru.nl}
\affiliation{Institute for Molecules and Materials, Radboud University, Heijndaalseweg 135, 6525 AJ, Nijmegen, The Netherlands}
\author{Alessandro Principi}
\email{a.principi@science.ru.nl}
\affiliation{Institute for Molecules and Materials, Radboud University, Heijndaalseweg 135, 6525 AJ, Nijmegen, The Netherlands}
\author{Mikhail I. Katsnelson}
\email{m.katsnelson@science.ru.nl}
\affiliation{Institute for Molecules and Materials, Radboud University, Heijndaalseweg 135, 6525 AJ, Nijmegen, The Netherlands}
\begin{abstract}
We address the question of the stability of the (fractional) quantum Hall effect (QHE) in presence of pseudomagnetic disorder
generated by mechanical deformations of a graphene sheet. Neglecting the potential disorder and taking into account only strain-induced random pseudomagnetic fields, it is possible to write down a Laughlin-like trial ground-state wave function explicitly. Exploiting the Laughlin plasma analogy, we demonstrate that in the case of fluctuating pseudomagnetic fluxes of relatively small amplitude both the integer and fractional quantum Hall effects are always stable upon the deformations. By contrast, in the case of bubble-induced pseudomagnetic fields in graphene on a substrate (a small number of large fluxes) the disorder can be strong enough to cause a glass transition in the corresponding classical Coulomb plasma, resulting in the destruction of fractional quantum Hall regime and in a quantum phase transition to a non-ergodic state of the lowest Landau level.
\end{abstract}
%
%
\maketitle

{\it Introduction}---Massless Dirac fermions were discovered \cite{massless_dirac1,massless_dirac2} in graphene via the observation of an unusual (``half-integer'') quantum Hall
effect (QHE) \cite{massless_dirac1,massless_dirac2,r1,r2,r3,Goerbig,Katsnelson:book} which is a manifestation of the existence of a topologically protected zero-energy
Landau level \cite{massless_dirac1,r2,Katsnelson:book}. This means that this level is not broaden by any inhomogeneity of the magnetic field. It was realized very soon after this
discovery \cite{Morozov:2006} that inhomogeneities of the effective magnetic field are unavoidable in graphene, due to the effect of pseudomagnetic fields induced by strain
(for a review, see Refs. \cite{Katsnelson:book,Vozmediano1,Vozmediano2}). In earlier works \cite{Morozov:2006,KG2008} random pseudomagnetic fields created by defects (such as intrinsic
and extrinsic ripples) were considered. Later it was theoretically predicted \cite{GKG1,GKG2} and experimentally confirmed \cite{Levy:2010} that (pseudo) Landau level quantization and valley
quantum Hall effect can be created in graphene by external smooth deformation with a trigonal symmetry, and that effective fields as high as hundreds of Tesla may be easily reached in this way
(an order of magnitude
stronger than what may be observed in conventional high-field magnetic laboratories).

The fractional quantum Hall effect (FQHE) has been experimentally discovered in graphene and its observation reported in Ref.~\cite{Du:2009,Bolotin:2009}. A very natural and interesting question is whether this state
is also protected, to some extent, with respect to inhomogeneities of the (pseudo)magnetic field or not. Here we answer this question within a framework of a model with random
pseudomagnetic fields but assuming the absence of potential disorder. In terms of deformations this means strong shear deformations and no dilatation \cite{Katsnelson:book}.
It was shown recently~\cite{Morpurgo} that at least in some graphene samples random strain-induced pseudomagnetic fields is indeed the main source of electron scattering and therefore the
model may be quite realistic.

{\it The model of pseudomagnetic disorder}---We neglect intervalley scattering and model graphene as two independent massless-Dirac-fermion systems, associated to the valleys ${\bm K}$ and ${\bm K}'$ at the corner of its hexagonal Brillouin zone. The two valleys differ for the sign of the strain-induced pseudomagnetic field they experience. A trial wave function for the zero-energy Landau level of a system of massless Dirac fermions in the presence of magnetic disorder can be constructed using the Aharonov-Casher solution \cite{Aharonov:1979}. We will briefly recall this construction. Consider a single electron in the presence of an arbitrary static magnetic
field in the direction orthogonal to the graphene plane, ${\bm B}({\bm r}) = B({\bm r}){\hat {\bm z}}$, and with support over a finite region of the two-dimensional (2D) space [here ${\bm r} = (x,y)$ is a 2D vector]. For given spin and valley quantum numbers, its Hamiltonian reads
\begin{equation}
 H = v{\bm \sigma}\cdot \left[-i\hbar {\bm \nabla} -\frac{e}{c}{\bm A}({\bm r})\right]
\end{equation}
where ${\bm \sigma}=(\sigma_x,\sigma_y)$ are Pauli matrices, $v \approx c/300$ is the Fermi velocity (here $c$ is the speed of light in vacuum), and ${\bm \nabla}\times {\bm A}({\bm r}) = {\bm B}({\bm r})$\cite{r3,Katsnelson:book,Goerbig}. In the Coulomb gauge [$\nabla {\bm A}({\bm r}) = 0$], introducing a ``magnetic scalar potential'' $\phi({\bm r})$ we rewrite
$ A_j({\bm r}) = - \partial_j \phi({\bm r})$,
where $j=x,y$ denotes spatial directions. The magnetic scalar potential obeys the equation
\begin{equation} \label{eq:harmonic_phi}
\nabla^2 \phi({\bm r}) = B({\bm r})
~.
\end{equation}
The off-diagonal component of the matrix Dirac equation for an electron in the zero-energy Landau level thus becomes
\begin{equation} \label{eq:Dirac_10_zero_energy}
 \left( \partial_x + i\partial_y+
 \frac{ie}{\hbar c}\partial_x \phi({\bm r}) +\frac{e}{\hbar c}\partial_y \phi({\bm r}) \right)\psi({\bm r}) = 0.
\end{equation}
where $\psi$ is one of the two components of Dirac spinor. We remind the reader that the second component is exactly zero, since states in the zero-energy Landau level are fully pseudospin-polarized. Substituting the {\it Ansatz}
\begin{equation}
 \psi({\bm r}) = e^{-\frac{e}{\hbar c} \phi({\bm r})} f({\bm r}),
\end{equation}
Eq.~(\ref{eq:Dirac_10_zero_energy}) reduces to
$
 \left(\partial_x + i\partial_y \right) f({\bm r}) =0,
$
whose linearly independent solutions are simply the polynomials
$
 f({\bm r}) = (x+iy)^\alpha
$
with $\alpha$ a positive integer.
The magnetic potential $\phi({\bm r})$ satisfies the Poisson-like Eq.~(\ref{eq:harmonic_phi}) and can therefore be written in terms of the 2D Green's function ($\propto \ln(r)$) as
\begin{equation}
 \phi({\bm r}) = \frac{1}{2\pi}\int d{\bm r}^\prime B({\bm r}^\prime) \ln \left(\frac{|{\bm r}-{\bm r}^\prime|}{r_0}\right)
 ~.
\end{equation}
Here $r_0$ is an arbitrary length scale. For our purposes, it is sufficient to study the case of a magnetic flux localized within a small region. From now on we will use the complex notation to denote the particle and flux position, {\it i.e.} given ${\bm r} = (x,y)$ we define the complex position $z=x+i y$. Then a
spin-polarized electron in the presence of a single flux located at $z_0 = x_0 + i y_0$ is described (up to a normalization constant) by the following wavefunction
\begin{equation}
 \psi(z)=|z-z_0|^{-\frac{\Phi}{\Phi_0}}z^\alpha \label{eq:SingleAC}
\end{equation}
for distances bigger than the characteristic flux size $r_0$, {\it i.e.} ${|z-z_0|\gg r_0}$.
Here $\Phi$ is the total value of the flux, $\Phi_0=\frac{2\pi\hbar c}{e}$ is the magnetic flux quantum, and $\alpha = 0, 1, \ldots$ parametrizes the angular momentum
of the particle ($\alpha$ is constrained to be smaller than $\Phi/\Phi_0$ for the wave function to have a proper behavior at infinity).
The effect of different fluxes on the single-particle wave function is clearly multiplicative.

The single-particle solution of \eqref{eq:SingleAC} makes it clear how to incorporate quenched (pseudo)magnetic fluctuations into the structure of the many-body Laughlin-like wave function \cite{Laughlin:1983}. The quenched (pseudo)magnetic field $\delta {\bm B}({\bm r})$ can be decomposed as a set of random fluxes situated at the points $\tilde{z}_j$, $j=1,\ldots, N_{\Phi}$ superimposed to the uniform magnetic field ${\bm B}_0 = B_0{\hat {\bm z}}$ which is responsible for the formation of the fractional state. In what follows we will take the magnetic length
$l_B=\sqrt{\hbar c/eB_0}$ to be the unit of length.
Hence,
\begin{equation}
 \Psi(z_1,\dots z_N)=\prod\limits_{i}^N\prod\limits_{j}^{N_\Phi} |z_i - \tilde{z}_j |^{M_j} \prod\limits_{k<l}^N (z_k - z_l)^m e^{-\frac14 \sum\limits_n |z_n|^2}
 ~, \label{eq:LLWF}
\end{equation}
where $m$ is an odd positive integer.
Any continuum distribution of the (pseudo)magnetic field can be described in this way in the limit $\Phi_j \rightarrow 0,N_\Phi \rightarrow \infty$.

The square of this wave function can be viewed as the partition function of a 2D Coulomb plasma moving on a background of randomly distributed (quenched) point charges, {\it i.e.}
${\cal Z} = \int dz_1\dots dz_N e^{- {\cal H}/m}$, where
\begin{eqnarray}
   {\cal H}&=&-2m^2\sum\limits_{k<l}^N \ln|z_k-z_l|+\frac{m}{2}\sum\limits_n^N |z_n|^2
   \nonumber\\
   &+&
   2m\sum\limits_i^N\sum\limits_j^{N_\Phi}\frac{\Phi_j}{\Phi_0}\ln|z_i-\tilde{z}_j|
   \end{eqnarray}


It was shown experimentally \cite{Morpurgo} that in many cases the sources of pseudomagnetic field can be considered as randomly distributed centers of deformations,
in a qualitative agreement with the ripple model \cite{KG2008}.
The typical size of these centers is much smaller than the distances between them, so we can effectively
treat them as point-like objects. The quenched pseudomagnetic disorder therefore emerges in a form of a set of
highly localized fluxes, and our problem reduces to the study of properties of a correlated electron gas
in presence of a random flux distribution coexisting with a homogeneous background magnetic field.

A disorder of a similar geometrical structure can be observed in graphene sheet put on a substrate, e.g. platinum surface \cite{Levy:2010}.
When epitaxial graphene is grown on such a substrate, bubbles of characteristic width about $10 \mbox{nm}$ and height $2\mbox{nm}$ tend to form.
In this case, the number of fluxes is much smaller and the value of each flux is much larger than in the case of rippled graphene.
For the sake of simplicity we assume that all pseudomagnetic fluxes have the same value $\Phi_j=\Phi$.

A few comments are now in order. While the two aforementioned types of disorder can be described by the same formal model,
they are pretty different on a phenomenological level. First of all, inhomogeneities of pseudomagnetic field in rippled graphene are normally not very strong,
with deviations $|\delta {\bm B}| \sim 1\mbox{T}$ on a length scale of $l \sim 1 \mbox{nm}$ \cite{Morozov:2006}, resulting in very moderate flux values $\Phi\simeq 10^{-3} \Phi_0$.
In contrast, nanobubbles observed in graphene on a metal substrate lead to very strong pseudomagnetic fields, of the order of $|\delta {\bm B}| \sim 300\mbox{T}$, localized within regions of characteristic size
$\sim (5-10) \mbox{nm}$ \cite{Levy:2010}. Hence, the corresponding fluxes are of the order of $\Phi \sim 10-15 \Phi_0$. Another fundamental difference is that out-of-plane deformations of
freely suspended graphene (ripples) can occur in both direction, hence the signs of $\Phi_j$ fluxes are randomly distributed, and each of them acts as local repulsive or attractive potential on the particles of the associated classical plasma. On the other hand, the fluxes due to nanobubbles are all of the same sign,
making the potential landscape for the Laughlin plasma either purely repulsive or purely attractive (depending on the relative sign of the background magnetic field and of the
fluxes). In what follows we mainly perform computations for the second case and demonstrate that a transition of the liquid plasma to a glass state is possible.
The first case turns out to be more trivial due to the weakness of the disorder, and it will be clear that both integer and quantum Hall effects are insensitive to deformations
of this kind.
\begin{figure}[t]
\begin{center}
\includegraphics[width=0.99\columnwidth]{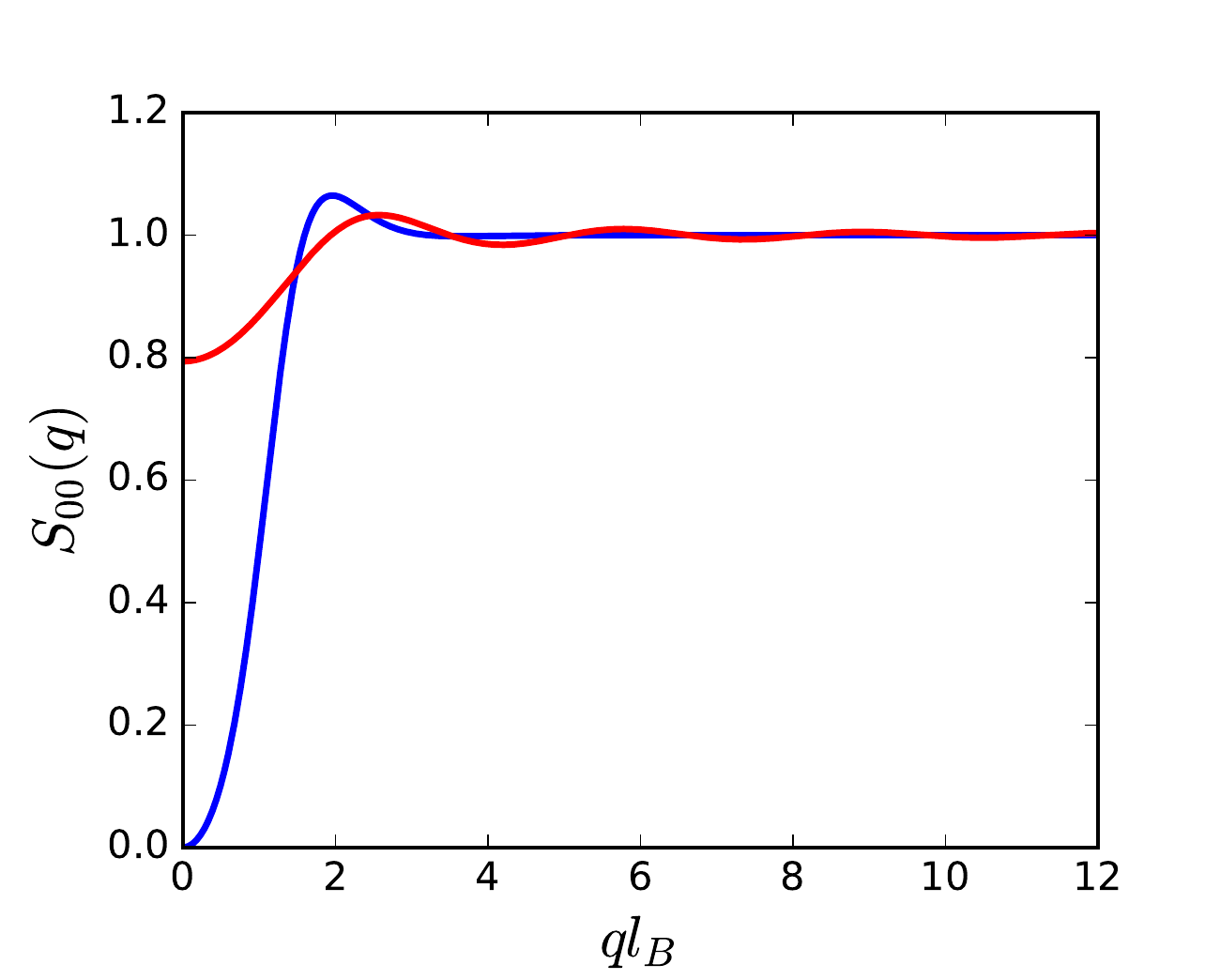}
\end{center}
\caption{The static structure factor $S_{00}(q)$ of short-range (blue) and long-range (red) correlated pseudomagnetic disorder at packing ratio $\eta=0.07$.
\label{fig:Sq00}}
\end{figure}

{\it Discussion of the model}---Great care should be taken here. Since charge carriers belong to two different valleys \cite{Katsnelson:book}, the same bubble deformation of the graphene
sheet induces a flux co-directed with the background magnetic field in one valley
(which result in an attractive potential for the classical plasma), and oppositely directed in the other valley. In the latter case it enters as a quenched repulsive potential in the action of the classical plasma. Since we allow for strong variations of (pseudo)magnetic field, to stay within the regime of validity of the model
we need to make sure that these inhomogeneities never lead to mixing between
Landau levels (LLs). In the valley where fluxes are co-directed with the background field the problem is not expected to occur: while the lowest Landau level (LLL) is protected upon variations
of the magnetic field, the gap between it and the next LL is bounded from below by the corresponding homogeneous value.
In the other valley the situation is potentially more dangerous, as the presence of oppositely directed large magnetic fluxes may imply the existence of zero-magnetic
field lines in the sample and, possibly, regions where LL merge. However, since the corresponding regions act on classical particles in the Laughlin plasma
as strong repulsive potentials, on the quantum level we expect the system to avoid occupying states within these domains, and the effect of level mixing should be mild.

Another potentially problematic aspect related to the existence of $B=0$ lines is the possibility of percolation through the bulk of the sample that destroys the quantization of the Hall conductivity \cite{Mirlin:1998}. But since we consider only the case of small packing ratio $\eta \equiv \rho_0 l_B^2$, and the distance between any two fluxes is much bigger than their characteristic size, such a line is an isolated loop circumventing a flux, and there is no chance of percolation.

\begin{figure}[t]
\includegraphics[width=0.99\columnwidth]{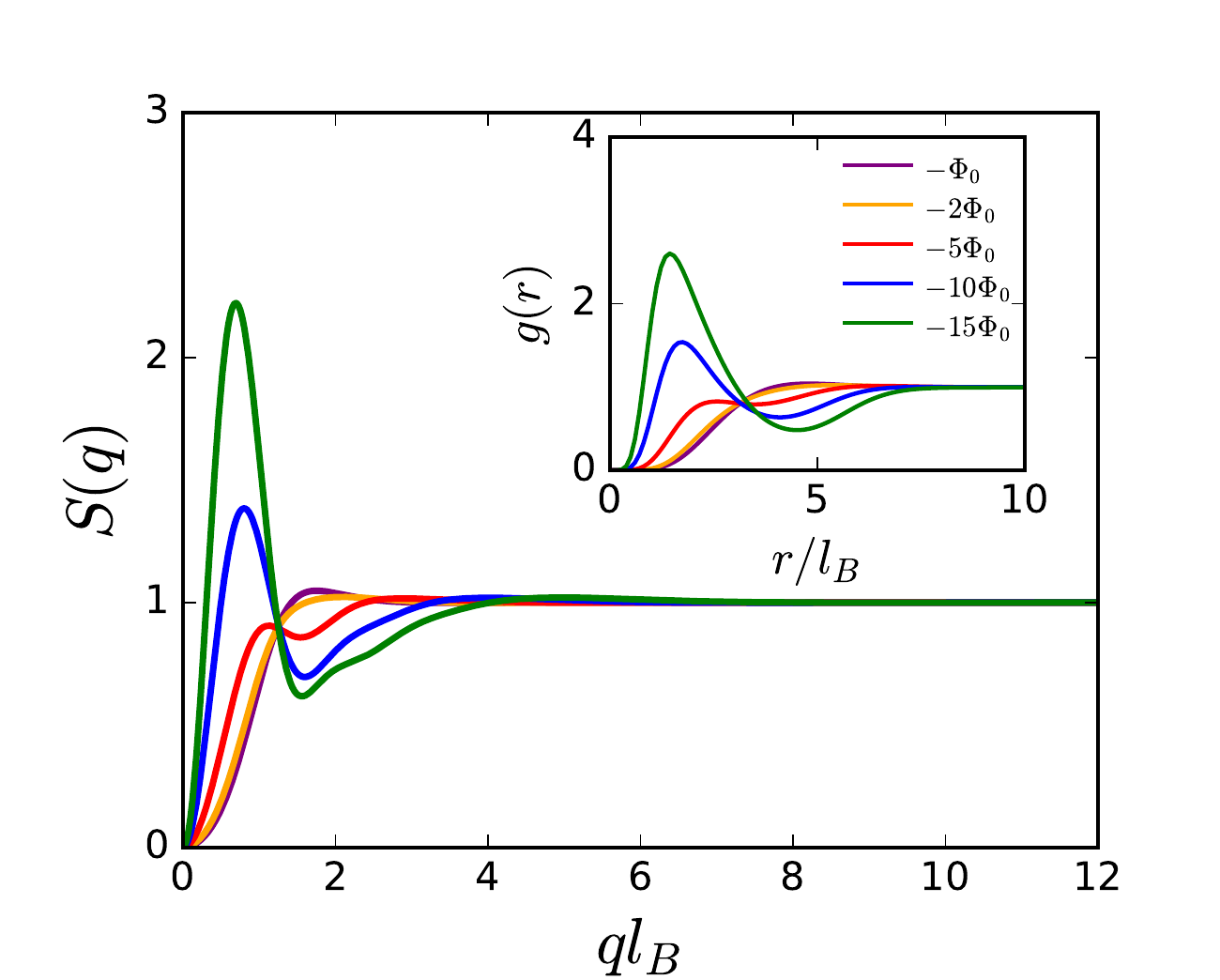}
\includegraphics[width=0.99\columnwidth]{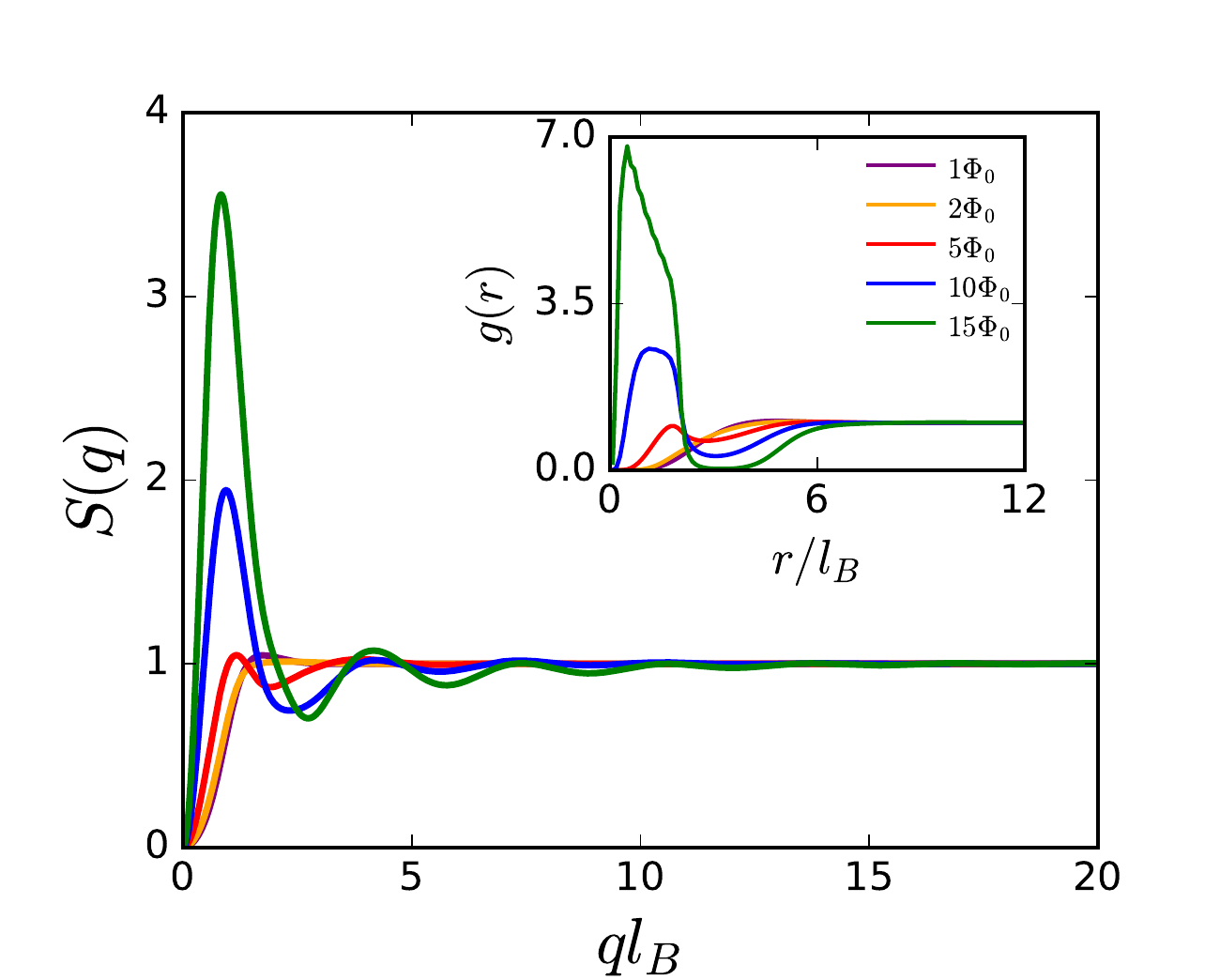}
\caption{Static structure factors $S(q)$ and pair correlation functions $g(r)$ (insets) of Laughlin plasma in presence of short-range correlated magnetic disorder: the cases of repulsive (top) and attractive (bottom) flux potentials.}
\label{fig:SqIncompressible}
\end{figure}

{\it Plasma static structure factor}---Having defined the model, we can numerically calculate the static structure factor of the Coulomb plasma, in the presence
of a number of static (disorder) charges, by means of the replica Ornstein-Zernike equations derived for
a partly quenched two-component fluid in Ref.~\cite{Given:1994}. Referring the reader to the original paper for a detailed discussion, hereafter
we just briefly quote the idea. In this language, the potential landscape provided by pseudomagnetic fluxes can be implemented as a frozen ``liquid''
whose direct ($c_{00}$) and full ($h_{00}$) pair correlation functions (and thus the static structure factor $S_{00}=1+\rho_0 h_{00}$) are fixed.
The correlation functions of the annealed component
can be obtained by solving the replicated system of equations
\begin{gather}
h_{01}=c_{01}+\rho_0 c_{00}\otimes h_{01}+\rho_1 c_{01}\otimes (h_{11}-h_{12}), \label{eq:ROZ}\\
h_{11}=c_{11}+\rho_0 c_{01}\otimes h_{01}+\rho_1 c_{11}\otimes h_{11} - \rho_1 c_{12}\otimes h_{12},\nonumber \\
h_{12}=c_{12}+\rho_0 c_{01}\otimes h_{01}+\rho_1 c_{11}\otimes h_{12} + \nonumber \\ \rho_1 c_{12}\otimes h_{11}-2\rho_1 c_{12}\otimes h_{12}, \nonumber
\end{gather}
where the $\otimes$ symbol stands for convolution $f\otimes g = \int f({\bm r}-{\bm r'})g({\bm r'})d^2{\bm r'}$.
Here the index $0$ denotes the quenched component ({\it i.e.} the magnetic fluxes), while $1$ and $2$ refer to two replicas of the annealed one (the electron liquid).
These equations are to be supplemented by
the closure relations. We use the hypernetted chain closure \cite{Laughlin:1983, Balescu:book}, which has been proven to give very
accurate results for quantum Hall plasmas
\begin{equation}
h_{ij}(r)=\exp\left( h_{ij}(r)-c_{ij}(r) - \beta v_{ij}(r)\right)-1,
\end{equation}
where $\beta v_{ij}(r)$ are radially symmetric interaction potentials ($\beta=m$ is the inverse temperature of the classical plasma), and replicas are required not to interact with each other directly, {\it i.e.} $v_{12}(r)=0$.
Here $\rho_0$ is the density of pseudomagnetic fluxes which can be estimated as follows. The typical background magnetic field which is required to develop a $\nu=\frac{1}{3}$ FQHE state is
$B_0\simeq 15\mbox{T}$ \cite{Bolotin:2009}, hence $l_B\simeq 6\, \mbox{nm}$. In experiments on graphene on a platinum substrate, a density of nanobubbles of about
$5$ per $2500\,\mbox{nm}^2$ \cite{Levy:2010} was observed, which in rescaled units would correspond to $\rho_0\simeq 0.07 \,{l_B}^{-2}$.
The particle density of itinerant electrons is instead fixed to
$\rho_1=1/(2\pi l_B^2 m)$ in a Laughlin state with the corresponding filling factor $\nu=1/m$.
Hereafter we set the magnetic length $l_B=1$ for convenience.

Before proceeding with solving \eqref{eq:ROZ} we have to fix the static structure factor $S_{00}(q)$ of the pseudomagnetic disorder.
While it is not known at all momentum scales, there are two qualitatively distinct cases that correspond to different behavior at small wave vectors. Inhomogeneities of the
pseudomagnetic field can be long- or short-range correlated. In the former case the correlator of pseudomagnetic vector potential $\langle|{\bm A}_{\bm q}|^2 \rangle$
behaves as $1/q^2$ \cite{Morpurgo,KG2008} for $q\to 0$, resulting in the correlator of pseudomagnetic fields
$S_{00}(q) \propto q^2 \langle|{\bm A}_{\bm q}|^2\rangle$ to approach a non-vanishing constant at $q=0$.
This kind of pseudomagnetic disorder is relevant for ripple-scattering dominated electronic transport \cite{Katsnelson:book,Morpurgo,KG2008}.
At the same time, short-range disorder with $\langle|{\bm A}_{\bm q}|^2\rangle$ approaching a constant at $q \rightarrow 0$ should always exist but does not lead to any appreciable contribution to
the electron mobility at zero magnetic field. However, as we will see in what follows, it can substantially effect the FQHE. In this case $S_{00}(q)\propto q^2$ at $q \rightarrow 0$.

A natural way to generate such model structure factors is to imagine for a second that, prior
to being quenched, the fluxes themselves were released to anneal as a liquid. If we assume that they interact with each other via a two-dimensional Coulomb potential,
we will end up with a structure factor corresponding to short-range disorder (represented by blue curve in Fig. \ref{fig:Sq00}).
If the ``annealing'' potential is taken to be of the hard-sphere type, we instead obtain a model of long-range disorder (red curve in Fig. \ref{fig:Sq00}).
Since real correlation functions of pseudomagnetic fields are unknown (and may be very different for different samples and substrates) we use those obtained from these two models.
This is enough to demonstrate a qualitative difference between long-range and short-range correlated disorder.

\begin{figure}[t]
\includegraphics[width=0.99\columnwidth]{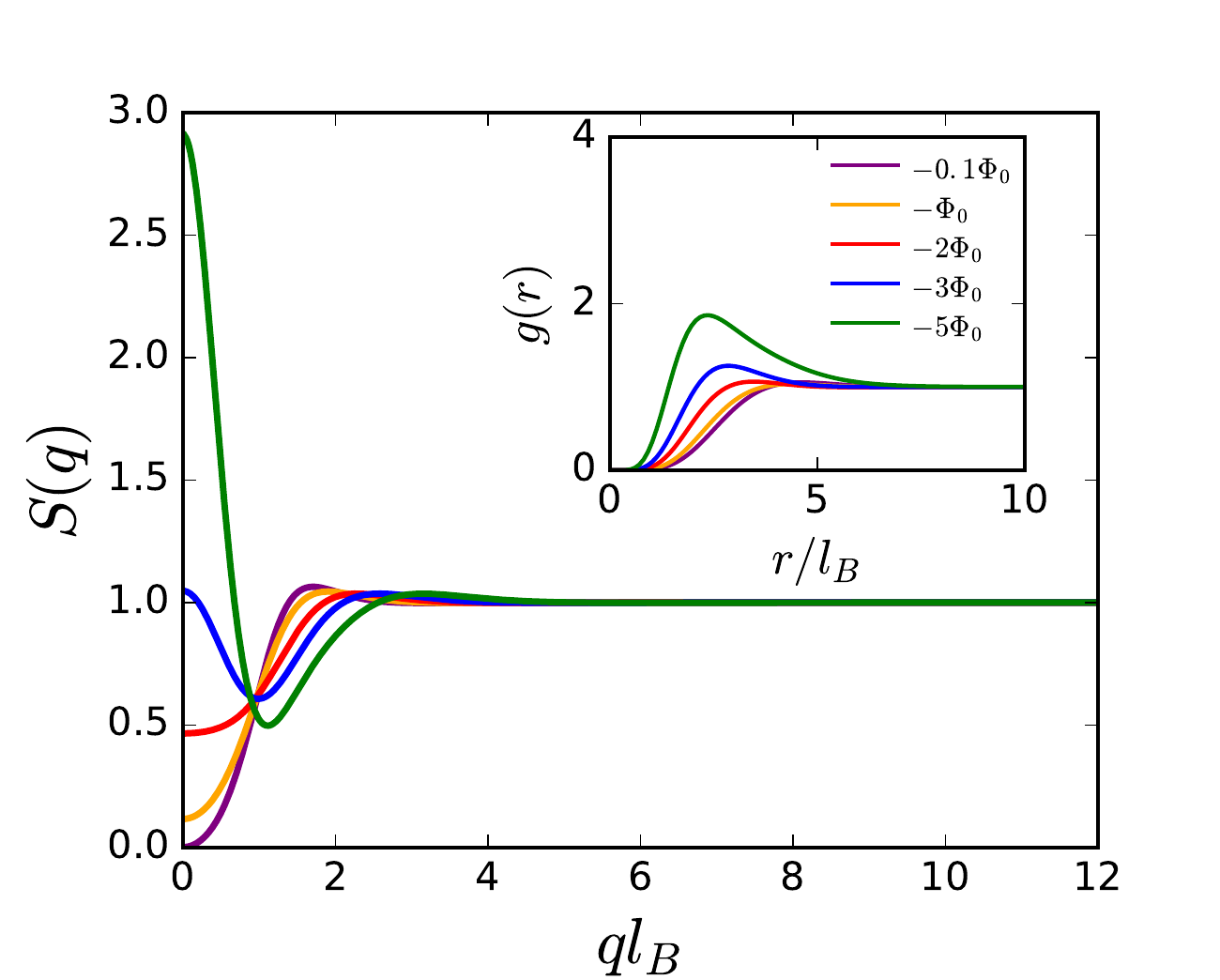}
\includegraphics[width=0.99\columnwidth]{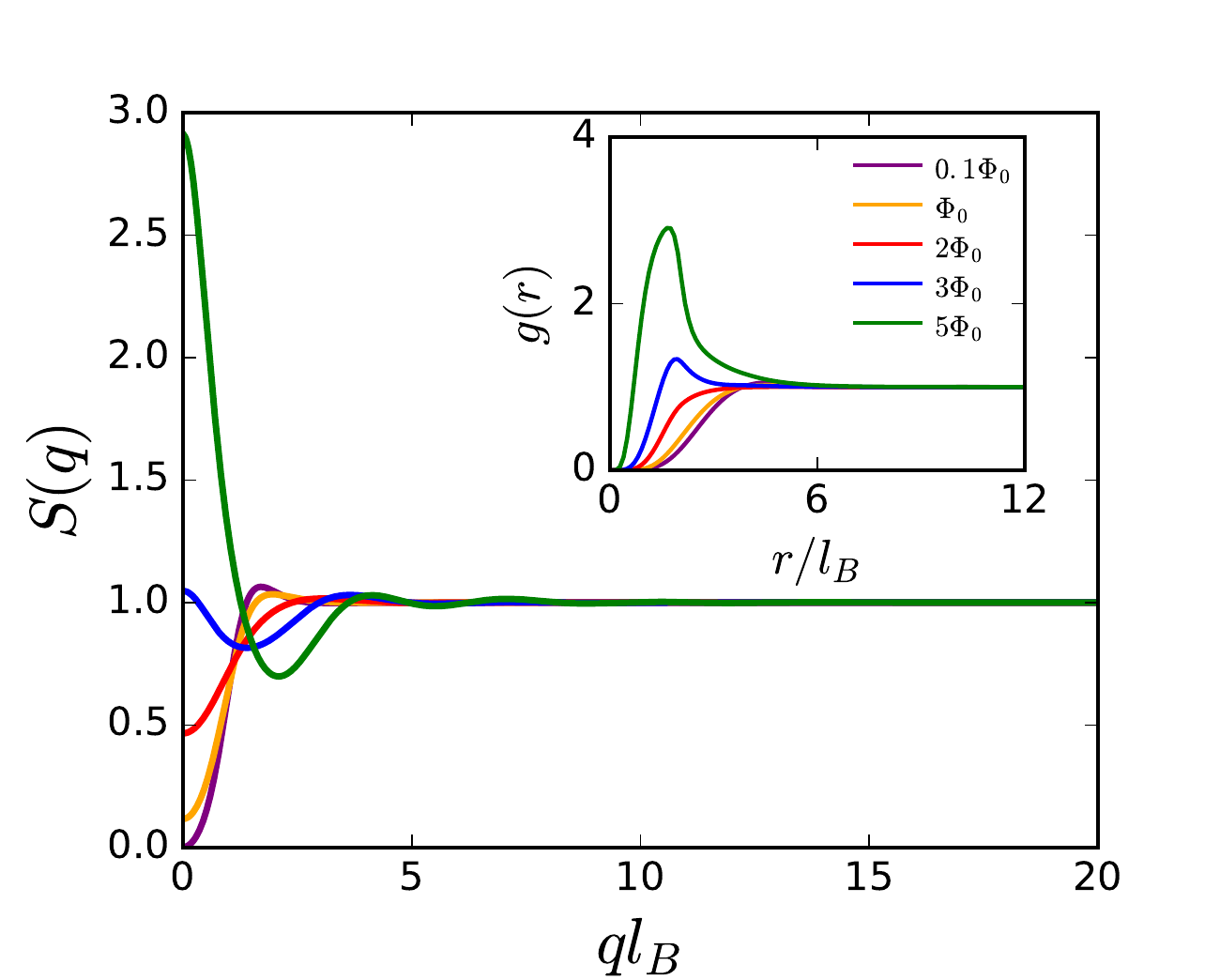}
\caption{Static structure factors $S(q)$ and pair correlation functions $g(r)$ (insets) of Laughlin plasma in presence of long-range correlated magnetic disorder: the cases of repulsive (top) and attractive (bottom) flux potentials.}
\label{fig:SqCompressible}
\end{figure}

The solutions of the replica Ornstein-Zernike equations in the two aforementioned cases are shown in Figs. \ref{fig:SqIncompressible} and \ref{fig:SqCompressible}, respectively.
We can see that the two types of pseudomagnetic disorder lead to very different physical effects. The structure factor $S_{11}(q)$ of a quantum Hall plasma in the presence of a short-range
disorder remains vanishing at $q\rightarrow 0$ whatever the strength of pseudomagnetic fluxes is. We can also check that the incompressibility sum rule
is always satisfied: $\frac{1}{2\pi}\int (g(r)-1) d^2r =-1$ \cite{Prange:book}.
On the other hand, long-range correlations in the pseudomagnetic disorder already at small strength $\Phi$ lead to a change in the small-$q$ behavior of the QHE plasma structure factor, making it compressible
and thus destroying the quantum Hall effect.

In the next section we will show that the high peak that emerges in $S_{11}(q)$ if the magnetic disorder is strong enough is a precursor of a glass phase transition
of the Laughlin plasma, which might be interpreted as a breakdown of ergodicity in the corresponding quantum ground state.

{\it Liquid-glass transition from mode coupling theory}---We use the mode-coupling theory \cite{Goetze_paper} to describe the liquid-glass transition.
It has been shown that this approach gives reliable results in proximity of the liquid-glass transition and that they agree with results obtained with the replica trick \cite{Goetze_Book}. Moreover, it has been recently shown \cite{Zamponi_paper,Zamponi_paper_2} that similar equations can be derived for a replicated hard-sphere
system, assuming that glassiness is self-induced and that the replica symmetry is spontaneously broken.
It has been shown that the mode-coupling approximation captures some of the
higher-order effects which are much more difficult to include in the replica formulation \cite{Zamponi_paper,Zamponi_paper_2}.

A glass is in general identified by the impossibility of relaxing local density fluctuations even in the long-time limit. It is indeed a disorder system that
gets (spontaneously or artificially) stuck in a certain configuration upon freezing, and does not relax to the equilibrium state as a consequence of the exponentially large
number of metastable minima. The idea of the mode-coupling theory is to directly calculate the density relaxation function,
$\phi(q,t) \equiv \langle \rho_{{\bm q}}(t) \rho_{-{\bm q}}\rangle/S_q$, and to show that it does not decay to zero in the limit of $t\to \infty$. Here $\langle\ldots\rangle$
is the average over the classical equilibrium distribution and $S(q) = \langle \rho_{{\bm q}} \rho_{-{\bm q}}\rangle/N$ is the static structure factor of the liquid.

In general, the Fourier transform of the density relaxation function, $\phi(q,\omega)$, can be rewritten in terms of the memory function $M(\omega)$ as
\begin{eqnarray}
\phi(q,\omega) = -\frac{1}{\omega - \Omega_q^2/[\omega + M(\omega)]}
~.
\end{eqnarray}
Here $\Omega_q = q^2 v_{\rm th}/S_q$, and $v_{\rm th} = k_{\rm B} T/m$ is the thermal velocity. The memory function is microscopically defined in terms of the
force-force correlation function, {\it i.e.} $\langle F_{\bm q}(t) F_{-{\bm q}} \rangle$, where the time evolution of the force fluctuation $F_{\bm q}(t)$ occurs in the
space orthogonal to the density and the longitudinal current (in the sense of the Zwanzig-Mori scalar product \cite{Goetze_Book,Zwanzig_paper,Mori_paper}). Essentially, the role of projection
is to freeze the dynamics of single-density fluctuations from the force-force correlator, treating them as static, and leave only the contribution of correlated fluctuations.

The memory function $M(t)$ is then split into two components, $M_{\rm reg}(t)$ and $M_{\rm MC}(t)$ \cite{Goetze_Book}. The former is obtained by projecting the dynamic
evolution in the space orthogonal to pair {\it density} fluctuations, {\it i.e.} treating them as frozen, while the latter contains only the contribution of these lowest-order
correlated fluctuations. It is then reasonably assumed that $M_{\rm reg}(t)$ does not contribute to the long-time dynamics of the system, and hence the main contribution to
glass formation comes from $M_{\rm MC}(t)$. The mode-coupling expression is then obtained by decoupling four-point density correlators in terms of two-point ones,
{\it i.e.} $\langle \rho_{{\bm q}_1}(t) \rho_{{\bm q}_2}(t) \rho_{{\bm q}_3} \rho_{{\bm q}_4}\rangle \simeq \langle \rho_{{\bm q}_1}(t) \rho_{{\bm q}_3}
\rangle \langle \rho_{{\bm q}_2}(t) \rho_{{\bm q}_4}\rangle$ (and permutations) \cite{Goetze_Book}. This is the fundamental step of the mode-coupling approximation,
by which correlated multi-pair fluctuations are expressed as convolutions (in momentum) of simple pair fluctuations. The convolution ensures the conservation of total momentum.

Finally, the mode-coupling contribution to the relaxation function in the limit $t\to \infty$ [${\bar \phi}(q) \equiv \lim_{t\to \infty} \phi(q,t)$] reads \cite{Goetze_Book,Goetze_paper}
\begin{eqnarray} \label{eq:MC_main}
\frac{{\bar \phi}(q)}{1 - {\bar \phi}(q)} &=& \frac{n S(q)}{2 q^2} \int \frac{d^2{\bm k}}{(2\pi)^2} \big| V({\bm q}, {\bm k}) \big|^2 S(k) S(|{\bm k}-{\bm q}|)
\nonumber\\
&\times&
{\bar \phi}(k) {\bar \phi}(|{\bm k}-{\bm q}|)
~,
\end{eqnarray}
where $V({\bm q}, {\bm k}) = {\bm q}\cdot{\bm k} c(k) +  {\bm q}\cdot({\bm k} -{\bm q}) c(|{\bm k} - {\bm q}|)$, and $c(k) = [1 - n S(k)]^{-1}$ is the direct
correlation function. Eq.~(\ref{eq:MC_main}) must be solved self-consistently. In the range of parameters where Eq.~(\ref{eq:MC_main}) has non-trivial solutions the
liquid can undergo a glass transition upon undercooling.

\begin{figure}[t]
\includegraphics[width=0.99\columnwidth]{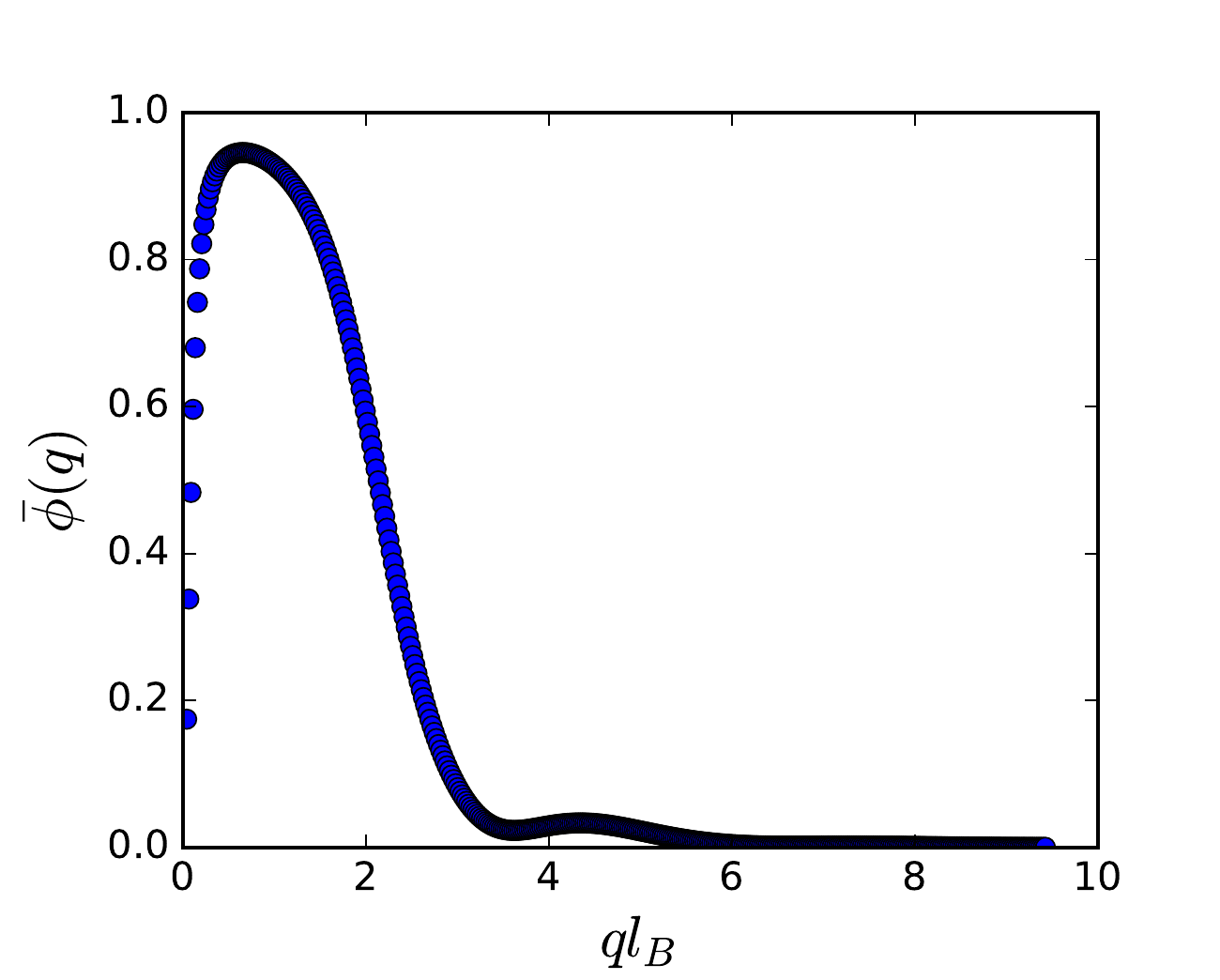}
\caption{Long-time relaxation function $\bar{\phi}(q)$ in the glassy phase (for packing ratio $\eta=0.07$ and flux strength $\Phi=15\Phi_0$).}
\label{fig:Relaxation}
\end{figure}

Using the structure factors computed in the previous section and plotted in Figs. \ref{fig:SqIncompressible} and \ref{fig:SqCompressible} as an input
for \eqref{eq:MC_main}, we can straightforwardly evaluate the late time relaxation function $\bar{\phi}(q)$ to see if the quantum Hall plasma exhibits a non-ergodic behavior.
An example of $\bar{\phi}(q)$ for a packing ratio $\eta = 0.07$ and flux value $\Phi=15\Phi_0$ (above the phase transition threshold)
is shown in Fig. \ref{fig:Relaxation}.

Having performed these calculation for single-sign pseudomagnetic disorder corresponding to graphene on a substrate, we can immediately conclude that
no glass formation can be expected in a model with smooth disorder. Although we can not exclude that the interplay of positive and negative magnetic fluxes can in principle enhance a bit the phase transition, the flux values are much smaller (by several orders of magnitude) than what is required for it to occur.

{\it Interpretation of the phase transition}---It is well known that inhomogeneous ground states can be realized in fractional quantum Hall systems at filling factors between those corresponding to incompressible states.~\cite{Efros_prb_1992,Koulakov_prl_1996,Fogler_prb_1996,Fogler_prb_1997,MacDonald_prb_2000,Spivak_prb_2003,Spivak_prb_2004,Goerbig_prb_2004} Striped or bubble phases have been predicted to have energies of the order of the homogenous (liquid) ground state, and to appear at intermediate filling factors whenever long-range interactions are present.~\cite{Efros_prb_1992,Spivak_prb_2003,Spivak_prb_2004} Striped phases are especially favorited by non-isotropic or dipolar-like interactions (like, e.g., those resulting from the screening of the Coulomb potential by nearby metal plates).~\cite{Efros_prb_1992} A characteristic hallmark of such phases is, for example, a non-homogeneous Hall conductivity.~\cite{Lilly_prl_1999,Cooper_prb_1999,Eisenstein_prl_2002,Fradkin_ann_rev_cmp_2010,Xia_nature_physics_2011} In clean systems and at low temperature such phases usually exhibit a long-range order.

An incompressible state in the presence of quenched magnetic disorder will exhibit a somewhat similar phenomenology. Local fluctuations of the filling factor will result in the formation of chaotic patterns. Although the pattern looks at a first sight completely chaotic, a careful study can reveal the hidden typical length scale, corresponding to a sharp maximum of the structure factor. Such patterns can, in general, be thought as stripe or bubble glass phases.~\cite{Schmalian_prl_2000,Principi_prb_2016} Their origin is due to the fact that the system finds itself frustrated by the presence of disorder, and wants to modulate with a period corresponding to the typical length scale but in all possible directions at the same time.~\cite{Schmalian_prl_2000,Principi_prb_2016}

We predict a phase transition as a function of the strength of the magnetic disorder. At all strengths the structure factor will exhibit a liquid-like behavior at small momentum ({\it i.e.} $S(q)\to q^2$ in the limit of $q\to 0$). A sharp peak develops at finite wavevectors, and grows with the strength of the quenched magnetic disorder $\Phi/\Phi_0$. This structure of $S(q)$ is crucial and leads to the emergence of a glassy behavior. The latter is revealed by the presence of non-decaying density fluctuations, encoded in the finiteness of the long-time part of the dynamical structure factor. Below the critical value of the strength of the quenched magnetic disorder, $S(q,t)\to 0$ at large time. However, above this critical value of $\Phi/\Phi_0$ it remains finite, signaling that density fluctuations do not decay over time. This behavior is typical for frozen systems.

We infer the presence of a phase transition from the behavior of the classical plasma associated to the deformed Laughlin wavefunction in the presence of quenched magnetic disorder. It is less clear what is the phenomenology of the quantum electron liquid above the critical value of the strength of the disorder. What seems to be clear is that the glass transition can be associated with a breakdown of the liquid-like behavior of the electron liquid, and therefore of the quantum Hall effect. This is analogous to the formation of a Wigner crystal at small filling fraction.
In our case the transition is driven not by the small density but by the presence of magnetic disorder and, depending on the strength of it, can occur at all filling fractions.

{\it Summary and conclusions}---In this paper we address the problem of the stability of a fractional quantum Hall state in the presence of pseudomagnetic disorder. Such a state find a natural realization in graphene subject to both a constant (large) magnetic field and strain fluctuations. Mechanical deformations of the graphene lattice can indeed generate disordered pseudomagnetic potentials with strengths up to hundreds of Teslas. To simplify our model we focused on this type of disorder, neglecting the potential contribution and the hybridization of Landau levels. In these conditions is it possible to write a trial ground-state by smoothly deforming the Laughlin wave function in the spirit of Aharonov and Casher~\cite{Aharonov:1979,Katsnelson:book}.

We treat this problem by means of the classical-plasma analogy. The square of the (deformed) Laughlin wave function is indeed formally identical to the action of a classical plasma, in which magnetic fluctuations play the role of quenched long-range (Coulomb) disorder. As in the classical theory of fluids, such a problem can be studied by numerically solving the replicated Ornstein-Zernike equations.~\cite{Given:1994} The disorder is represented as a quenched liquid with a given static structure factor, while the classical particles are treated as a fluid embedded into this disorder matrix. Two models of pseudomagnetic disorder are discussed, namely with long- and short-range correlations. Long-range correlated disorder occurs in the presence of ripples, and is characterized by structure factor which does not vanish in the limit $q\to 0$.~\cite{Katsnelson:book} Its effect on the fractional quantum Hall state turns out to be dramatic. Even for small concentrations of a weak disorder, the structure factor of the annealed component ({\it i.e.} the electron liquid) does not vanish in the limit of $q\to 0$. This in turn implies that the liquid becomes compressible and the fractional quantum Hall state is completely destroyed.~\cite{Laughlin:1983}

On the other hand the short-range disorder, whose quenched structure factor resemble that of a normal liquid, has a more subtle effect on the fractionalized state. Such disorder occurs, for example, in graphene on metals, where few bubbles due to the substrate can induce very localized and strong pseudomagnetic fluctuations.~\cite{Levy:2010} In this case, although the static structure factor resemble that of a liquid ({\it i.e.} with a $q^2$-like behavior at small momenta), it can develop strong peaks at finite wavevector. Such peaks can in turn induce a glass transition in the system,~\cite{Goetze_paper} which is signaled by the infinite stiffness of the system towards relaxing local density fluctuations. In this case the long-time structure factor, which normally vanishes in a liquid, remains finite. This phase transition occurring in the Coulomb plasma can be associated with a destruction of the fractional quantum Hall regime of the electron liquid, which enters into a non-ergodic state of the lowest Landau level.

{\it Acknowledgments}---The authors acknowledge support from the ERC Advanced Grant 338957 FEMTO/NANO and from the NWO via the Spinoza Prize.

\end{document}